# Untangling 3D atomic reconstruction in twisted bilayer 2D crystals via dark field transmission electron microscopy

Pankaj Kumar[1,2], Michel Bosman[1], Nikolai Lavrentev[1], He Zheng[1], Ding Peng[3], Kostya S. Novoselov[1,2,4], Tatiana Latychevskaia[3,*]

[1]Department of Materials Science and Engineering, National University of Singapore, Singapore, 117575, Singapore

[2]Institute for Functional Intelligent Materials, National University of Singapore, 117544 Singapore

[3]Paul Scherrer Institute, Forschungsstrasse 111, 5232 Villigen, Switzerland

[4]National Graphene Institute, University of Manchester, Oxford Road, Manchester, M13 9PL, UK

*Corresponding author: tatiana.latychevskaia@psi.ch

## Abstract

Reconstruction of the atomic crystal structure in twisted 2D materials has been demonstrated to be responsible for multiple exciting phenomena in van der Waals heterostructures, from the appearance of flat bands in twisted bilayer graphene to Wigner crystallization in transition metal dichalcogenides (TMDs). However, there are still no experimental methods for accessing the 3D atomic distributions nor models that describe the exact atomic shifts in such reconstructed structures, which significantly impedes the development of the field. Dark field (DF) transmission electron microscopy (TEM) has been conventionally employed to visualize the local in-plane atomic displacements. Here we expand this method to obtain a full description of the reconstructed atomic systems and demonstrate the quantitative relations between the local stacking and the intensity in the DF image. We show how local 3D atomic displacements and the interlayer distance can be extracted from a DF image.

## Keywords

2D crystals, graphene, transition metal dichalcogenides, twisted bilayer, electron microscopy, dark-field imaging

## Main text

In marginally twisted bilayer materials, the moiré structure creates new Bloch waves for the electrons in the crystals with moiré periodicity, which results in a strongly modified band structure and leads to new properties of the resulting material [1-7]. In these materials the atomic positions often get reconstructed – the atoms of the two layers interact with one another and settle to the relaxed positions that are different from the positions of the atoms in the non-interacting layers. Strong atomic reconstruction usually occurs when the moiré period is not very large, corresponding to a small (<1°) twist angle between the adjacent layers, because it allows to distribute the strain. The resulting system exhibits a variety of different local stackings at different positions within the structure. Van der Waals interaction tends to expand the most energetically favourable stackings in exchange of the less favourable. The process is limited by the elastic energy associated with such reconstructions.

The relaxed structures are usually simulated by means of density functional theory (DFT) and molecular dynamic (MD) methods [8-12]. However, in most structures of interest, the reconstructed areas include hundreds of thousands (or even more) of atoms, which is beyond the capabilities of modern DFT, and MD potentials are not always accurate enough to precisely capture the van der Waals interaction between different atoms.

In the realistic reconstructed van der Waals heterostructures, a well-defined stacking has been observed only exactly at the high-symmetry points, with the atomic displacements smoothly transitioning between these points. Quantitative knowledge of exact values of these atomic displacements is extremely important for the understanding of the electronic, optical, mechanical, and chemical properties of these systems. At the same time, it is extremely difficult to access these local 3D atomic distributions in such layered materials [13-15] either experimentally or theoretically. The stacking order in bilayer materials has been conventionally visualized by dark field (DF) TEM microscopy [16-20], and less conventionally by low-energy electron microscopy [21]. However, the previous DF TEM studies address only the relative 2D in-plane atomic positions in the layers. In this study, we demonstrate a framework for high-precision retrieval of the three-dimensional atomic coordinates in a relaxed twisted bilayer system from its DF image.

To make our approach most general, we will consider the case of crystals with broken inversion symmetry, such as hexagonal boron nitride (hBN) or 2D-$MoS_2$. The systems exist in two possible aligned stacking – parallel and antiparallel; the latter occurs most often in nature. 2D-$MoS_2$ structures are shown in Fig. 1a,c. For simplicity, in the rest of the paper we will focus on the parallel stacking, with antiparallel producing broadly similar results, and one can utilise our methodology to easily derive formulae for such a case.

For the case of parallel stacked bilayer 2D-$MoS_2$, the least favourable local stacking configuration is the XX(P), which is found in the vertices of the reconstructed moiré domains, Fig. 1a,b. The regions of MX and XM stacking tend to expand (gaining van der Waals energy), creating alternating regions of MX/XM domains arranged into a trigonal pattern [19]. The centres of XX(P) stacking are surrounded by the intermediate (IM) stackings. The MX/XM regions are separated by the topologically protected domain walls (DW) or saddle point (SP) stacking (Fig. 1b,d) [20, 22].

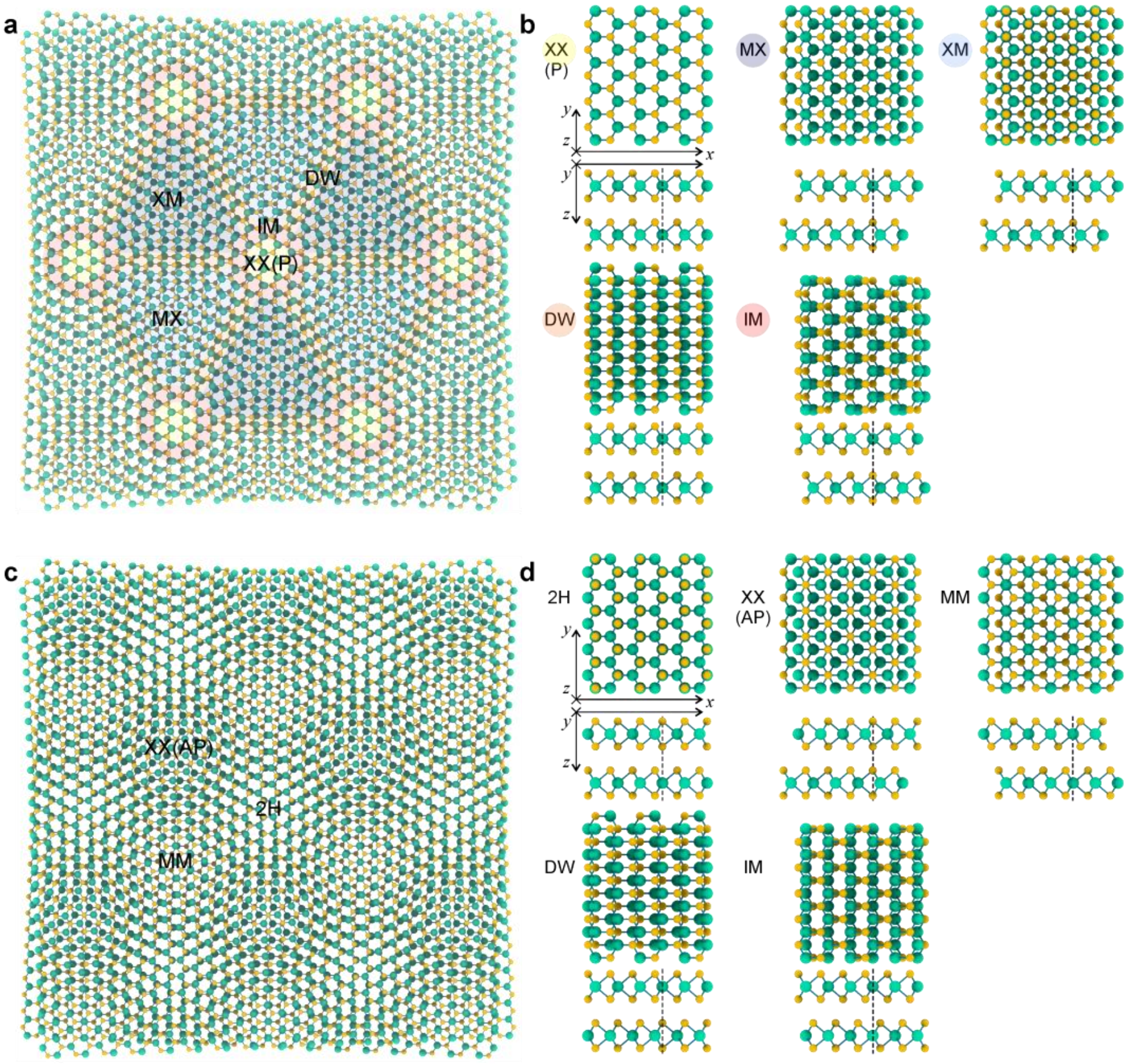

Fig. 1. Atomic stacking in a system of twisted honeycomb crystals without the inversion symmetry (shown for non-relaxed atomic coordinates). (a) – (b) Twisted parallel (3R) and (c) – (d) antiparallel (2H) bilayer 2D-$MoS_2$ and regions with different stackings.

Here we would like to suggest that DF TEM is ideally suited for the extraction of the atomic coordinates of the reconstructed twisted bilayer. In DF imaging mode, a diffraction peak (typically of a first or second order) is selected in the back focal plane (BFP) of the objective lens, Fig 2a. Experimental DF of a twisted 2D-$MoS_2$ is shown in Fig. 2b, exhibiting strong contrast between MX and XM regions; the corresponding bright field (BF) image is shown in Fig. 2c. BF image exhibits very weak intensity contrast between the regions of different stacking due to the strong direct beam. Although BF imaging is less common than DF imaging, we show below that adding BF can be necessary in order to solve some ambiguous atomic arrangements.

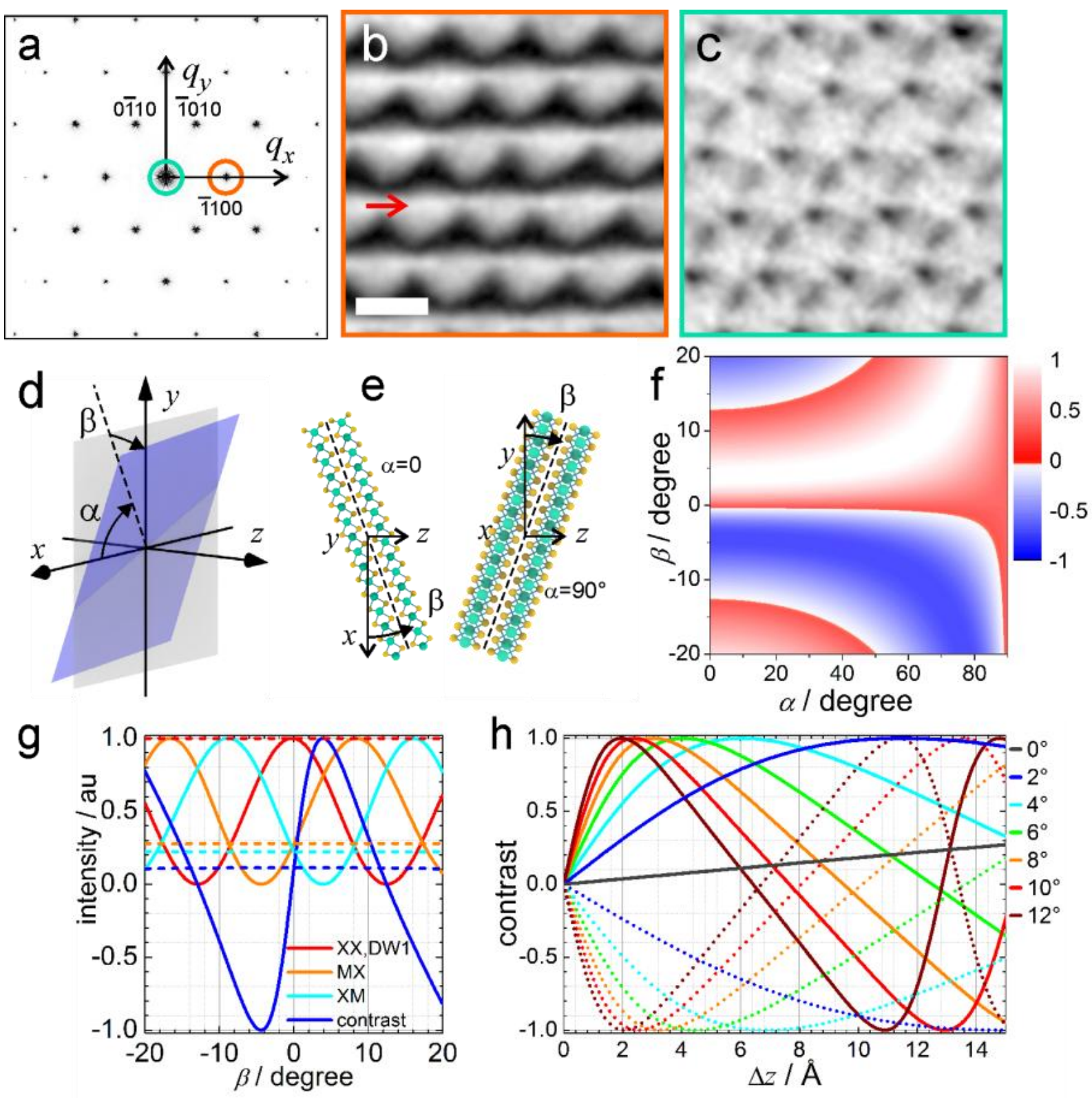

Fig. 2. Dark field (DF) transmission electron microscopy (TEM) of twisted 2D-$MoS_2$; $\bar{1}100$ peak is selected. (a) Intensity distribution in the back focal plane (BFP) of the objective lens; a diffraction peak is selected for dark field (DF) (orange circle), and the zero-order peak (cyan circle) – for bright field (BF) imaging. (b) and (c) experimental DF and BF images of a twisted 2D-$MoS_2$ structure, respectively; scalebar is 20 nm. (d) Schematics of sample tilt described by two angles $\alpha$ and $\beta$. (e) Illustration to the sample tilt by angle $\beta$ in the ($y$, $z$) plane ($\alpha$ = 90°) and in the ($x$, $z$) plane ($\alpha$ = 0). (f) MX/XM contrast as function of sample tilt $\alpha,\beta$ at $\Delta z$ = 6.2 Å. (g) Simulated intensities and the MX/XM contrast as a function of the tilt angle $\beta$ ($\alpha$ = 0) shown in solid lines, and $\beta$ ($\alpha$ = 90°) shown in dashed lines, respectively; $\Delta z$ = 6.2 Å. (h) MX/XM contrast as a function of the interlayer distance $\Delta z$ for a series of the sample's tilts in the ($x$, $z$) plane ($\beta$, $\alpha$ = 0), the dependencies for the tilts in negative direction $\beta < 0$ are shown in dotted lines.

The intensity of DF image of a bilayer system is described by

$$I^{(n)} = 2I_0\left[1+\cos\left(\boldsymbol{q}^{(n)}\Delta\boldsymbol{r}\right)\right], \quad (1)$$

where $\boldsymbol{q}^{(n)}$ is the momentum transfer vector corresponding to the diffraction peak of the *n*-th order selected in the BFP, and $\Delta\vec{r}$ is the vector describing the local 3D translational shift between the two layers (Supporting Note 1: Interlayer shifts in different stackings), and $I_0$ is the uniform intensity produced by a single layer. The two layers are assumed to be identical, where each layer can be composed of atoms with different scattering strength. Even in the absence of any lateral shift between the layers ($\Delta x = \Delta y = 0$), the interlayer distance $\Delta z$ creates variations in the local intensity contrast. The intensity is maximal for the regions where $\boldsymbol{q}^{(n)}\Delta\boldsymbol{r} = 0$, these are the XX regions and the domain walls (DW) regions where the interlayer shift $\Delta\boldsymbol{r}$ is orthogonal to the momentum transfer $\boldsymbol{q}^{(n)}$ (indicated by the red arrow in Fig. 2b). Thus, the selected diffraction peak can be directly identified from the orientation of the bright, intense line in DF image.

By selecting the direction of the *x*-axis in the sample plane so that $\Delta x_{\mathrm{MX,\,XM}} = \mp a_0$ (Fig. 1), where $a_0 = a/\sqrt{3}$ and *a* is the lattice constant, and assuming that a first-order diffraction peak is selected for DF imaging, the intensities of the MX and XM regions from Eq. (1) are:

$$I^{(1)}_{\mathrm{MX,XM}} = I_0\left\{1+\cos\left[q^{(1)}_x a_0 \pm \frac{4\pi\lambda}{3a^2}\Delta z\right]\right\}.$$

The interlayer distance can be determined from the contrast between the MX and XM regions, which does not require the knowledge of $I_0$:

$$C^{(n)}_{\mathrm{MX/XM}} = \frac{I^{(n)}_{\mathrm{MX}} - I^{(n)}_{\mathrm{XM}}}{I^{(n)}_{\mathrm{MX}} + I^{(n)}_{\mathrm{XM}}}.$$

For the peak $\bar{1}100$, $q_x^{(\bar{1}100)} = 4\pi/\sqrt{3}a$ and $q_y^{(-1100)} = 0$ (shown in Fig. 2a), the intensities are $I^{(\bar{1}100)}_{\mathrm{MX,XM}} = I_0\left\{1+\cos\left[\frac{4\pi}{3}\left(1\pm\frac{\lambda}{3a^2}\Delta z\right)\right]\right\}$, and the contrast is $C^{(\bar{1}100)}_{\mathrm{MX/XM}} \approx \frac{4\pi\lambda}{\sqrt{3}a^2}\Delta z.$

When a second-order diffraction peak is selected for DF imaging, $q_x^{(2)}\Delta x_{\mathrm{MX,\,XM}} = 0$ or $\pm 2\pi$, which gives $I^{(2)}_{\mathrm{MX}} = I^{(2)}_{\mathrm{XM}}$, and thus the contrast between the MX and XM regions is zero. This was previously observed experimentally [19, 23]. In the following, we will consider the situation when a first-order diffraction peak is selected for DF imaging.

In a DF TEM experiment, the sample is typically tilted by a few degrees to achieve the maximal contrast of the DF image. The sample can also be locally tilted due to the corrugations. Thus, the local tilt of the sample is not exactly known. For a tilted sample, the intensity and contrast in DF image are described by the same Eq. (1), but the coordinates of the 3D translation vector $\Delta\boldsymbol{r}$ are rotated according to the sample tilt. An arbitrary tilt can be described by two angles: $\alpha$ is the angle in the (*x*, *y*)-plane counted from the x-axis and indicating the axis along which the tilt is done, and $\beta$ is the angle of the tilt along that axis (Fig. 2d). When $\alpha = 0$ the sample is tilted in the (*x*, *z*)-plane, and when $\alpha = 90°$ the sample is tilted in the (*y*, *z*)-plane (Fig. 2e). The coordinates of the 3D translation vector $\Delta\boldsymbol{r}$ are transformed as $\Delta\mathbf{r}' = R_1(\alpha)R_2(\beta)R_1(-\alpha)\Delta\mathbf{r}$, where $R_1$ and $R_2$ are the rotation matrices in the (*x*, *y*) and (*x*, *z*) planes, respectively (Supporting Note 2: Sample tilt).

Using Eq. (1), the intensity in DF image at regions with different stacking can be calculated for an arbitrary tilt of the sample. Simulated intensities of XX, MX and XM regions and the contrast as a function of the sample's tilt and the interlayer distance are shown in Fig. 2f,g,h. These dependencies were cross-validated by simulating DF images of the bilayer atomic structures via multislice simulations (Supporting Note 3: Multislice simulations). From the simulations shown in Fig. 2g, it can be seen that tilting of the sample in the (*x*, *z*)-plane ($\alpha = 0$) results in a noticeable change of the intensity and contrast, while tilting the sample in the (*y*, *z*)-plane ($\alpha = 90°$) does not change either the intensity or the contrast of a DF image.

The obtained dependencies can help to find the optimal tilt angle for maximal contrast. For a twisted bilayer 2D-$MoS_2$ with interlayer distance $\Delta z$ = 6.2 Å, when tilting in the (*x*, *z*)-plane ($\alpha = 0$) the maximal contrast is achieved at 4.2° (Fig. 2g,h). For a TBG sample with interlayer distance $\Delta z$ = 3.35 Å, the tilt angle should be 6.07°, which agrees well with the previously reported experimental tilt of 5° [19].

Most simulations shown here were made for a parallel (3R) 2D-$MoS_2$ system; the same system was used in the experiment. Similar analysis can be applied for the antiparallel (2H) 2D-$MoS_2$ system (Supporting Note 4: Antiparallel (2H) 2D-$MoS_2$ system).

In this section, we present a simple methodology, which can be implemented for the extraction of atomic reconstruction for twisted bilayer samples from experimental DF images. We demonstrate that even realistic experimental samples, which typically contain defects, corrugations, etc. can be utilised for such purpose. We used a marginally twisted bilayer 2D-$MoS_2$ sample that was prepared and imaged as described in Supporting Note 5: Sample preparation and illustrated in Supporting Figure 2. Various fingerprint intensity patterns were observed in the DF image that can be assigned to the following regions in the sample: regions of alternating MX/XM stacking where characteristic "triangles" are observed in DF images [15], contrast inversion due to sample tilt, non-interacting layers, and strained layers. These structural arrangements are discussed in more detail below. The DF and BF images of the entire sample are provided in Supporting Figure 3.

Free-standing 2D crystals are typically corrugated, which can be seen directly in the DF image. For accurate extraction of atomic reconstruction, it is crucially important to account for such corrugations. Within a ripple, the sample's tilt can change its sign, which in turn changes the sign of the contrast (Fig. 2h). For 2D-$MoS_2$ bilayer system, the contrast changes at $\pm 13^{\circ}$ tilt (Fig. 2g,h). This change of the contrast's sign manifests itself in the experimental DF images as flipping of the MX/XM "triangles", Fig. 3a. The BF image is only slightly changed under the sample tilt, Fig. 3b. The simulated DF and BF images of a twisted (2°) 2D-$MoS_2$ bilayer system were obtained by using a multislice approach (Supporting Note 3: Multislice simulations) and are shown in Fig. 3c – h. For the simulations shown in Fig. 3, we used the following analytical model for relaxed coordinates. The relaxed structure was modelled by rotation of the atoms around the XX(P) nodes [24], by angle $\theta(r) = \pm\left[\frac{\theta_0}{2} + \Delta\theta(r)\right]$, with $\Delta\theta(r) = \frac{\theta_0}{2}\exp\left(-\frac{r^2}{2d^2}\right)$, where $\theta_0$ is the twist angle ($\theta_0 = 2°$), and $d$ = 3 nm; $d$ was determined by performing a series of simulations and selecting the results that best match the experimental DF and BF images. The twist angle $\theta_0 = 2°$ was selected to be much larger than it is required for atomic relaxation ($\theta_0$<1°) and than it was measured in the experiment ($\theta_0 =$0.6 – 0. 8°) in order to obtain a short period of the moiré structure and speed up the calculation time; this is also the explanation for the blurred appearances of the "triangles". Using this analytical model for relaxed coordinates, we were able to create a noticeable relaxation at a relatively large twist angle of 2° and with this reduce the size of the region exhibiting alternating stacking in the moiré superlattice, which required a reasonable computational time for multislice calculations.

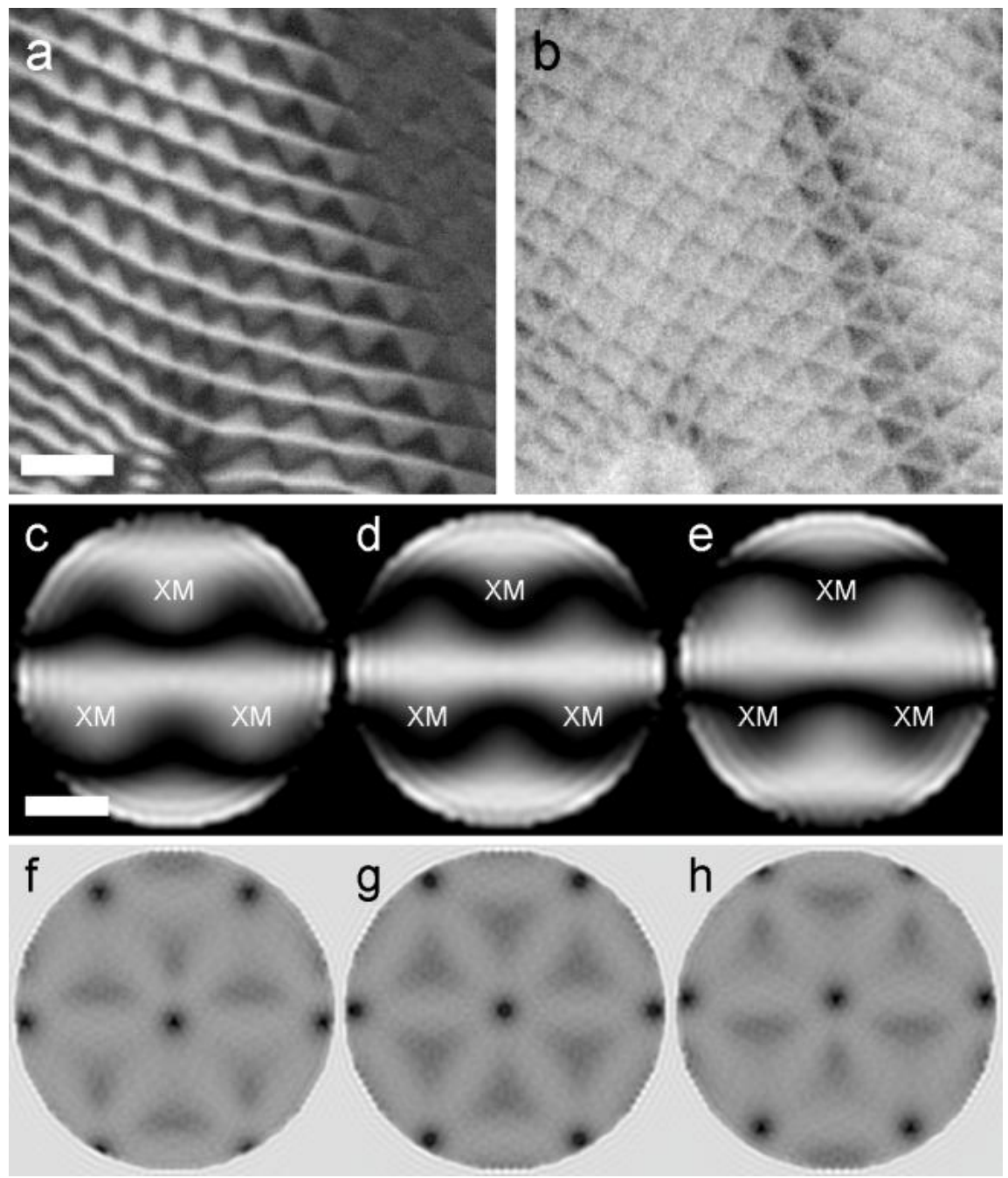

Fig. 3 DF image of corrugated twisted bilayer 2D-$MoS_2$ sample. (a) DF and (b) BF experimental images of MX/XM alternating regions ("triangles") with the inversion of the contrast across the sample; scalebar, 50 nm. (c) – (e) simulated DF and (f) – (h) BF images of a relaxed 2° twisted bilayer 2D-$MoS_2$ (interlayer distance 6.2 Å) tilted in the (*x*, *z*)-plane ($\alpha = 0$) by $\beta$ = -4° (c,f), 0° (d,g) and +4° (e,h); scalebar, 5 nm.

Interlayer distance can be evaluated from the MX/XM contrast as follows. For the experimental DF image shown in Fig. 2b, the twist angle evaluated from the period of the "triangles" pattern is 0.9°. Assuming that $\beta < 0$, and using the results shown in Fig. 2g,h, we assign the bright regions to XM and dark regions to MX stackings. The evaluated contrast is $C_{MX/XM}$ = -0.69, which gives the range of possible interlayer distances and tilts (Fig. 2h): $\Delta z$ = 6.2 Å at $\beta$ = -2.08° to $\Delta z$ = 8 Å at $\beta$ = -6.2°.

The local interlayer shifts can be retrieved from the DF image using the following considerations. For two layers twisted by $\pm\vartheta_0/2$, the local translational shift between the two layers at a coordinate $(x, y)$ is given by: $\Delta x(x, y) = 2y\sin(\vartheta_0/2)$, $\Delta y(x, y) = -2x\sin(\vartheta_0/2)$. Thus, for a non-reconstructed (rigid) atomic coordinates, $\Delta x(x, y)$ is a monotone function of the *y*-coordinate and $\Delta y(x, y)$ – of the *x*-coordinate, respectively (Fig. 4a,b). In this rigid system, although the stacking order is changing along the *x*-axis, $\Delta x$ does not depend on the *x*-coordinate, and thus the intensity of DF image (determined by $q_x\Delta x$ product) is constant along the *x*-coordinate (Fig. 4c). In the DF image, the period of fringes along the *y*-axis is $d_m = \frac{a\sqrt{3}}{4\sin(\vartheta_0/2)}$, which is the distance between the planes in the moiré lattice. In a reconstructed (relaxed) system, on the other hand, $\Delta x(x, y)$ and $\Delta y(x, y)$ exhibit slightly perturbed monotone distributions (Fig. 4d,e), and the fringes in the DF image turn into "triangles" (Fig. 4f).

The fact that $\Delta x(x, y)$ is a monotone function of the *y*-coordinate has the following interesting consequences. It means that the argument of the cosine function $q_x \Delta x + q_z \Delta z$ (at $q_y = 0$) in Eq. (1), is a monotone function of the *y*-coordinate, even in the presence of the sample tilt. Therefore, the cosine function, and with this, the DF image as a function of the *y*-coordinate will exhibit a function that continuously goes through minima (zero) and maxima of intensity. The positions of the minima and maxima are illustrated by the yellow and red dashed lines in Fig. 4. This also explains the topologically protected states observed in twisted bilayer systems. In a twisted bilayer system, the location of highly-symmetrical stacking (AA points, AB, BA stackings, domain walls) can be assigned a certain specific atomic arrangement with certain specific $\Delta x(x, y)$. Should the structure be stretched or tilted, the same shift $\Delta x$ (and with this the same contrast in DF image) will always be present in the system but will be found at slightly different coordinates $(x', y')$. Thus, in the stressed/strained/tilted structures, the atomic positions can be different, but the number of points of the highly symmetrical stacking always remains the same as in the non-stressed sample.

The translational shifts $\Delta x(x, y)$ were retrieved from the experimental DF image by using Eq. (1) and are shown in Fig. 4h-i. The minima and maxima in the DF image along each *x*-coordinate were determined, and surfaces of minima $I_{\min}$ and maxima $I_{\max}$ were obtained by interpolation. A normalized DF was obtained as DF' = (DF − $I_{\min}$)/($I_{\max}$ − $I_{\min}$). Then, using Eq. (1), acos of (2DF'-1) was calculated, and the obtained phase $\Delta\varphi$ was unwrapped. The translational shifts $\Delta x(x, y)$ were extracted from $\Delta\varphi$ using relations $\Delta\varphi = q_x \Delta x' + q_z \Delta z'$, where $\Delta z' = \Delta x \sin\beta + \Delta z \cos\beta$, $\Delta x' = \Delta x \cos\beta - \Delta z \sin\beta$, and where $\beta$ = -2.08° at $\Delta z$ = 6.2 Å as evaluated from the experimental DF image. The linear shifts $\Delta x(x, y)$ due to twist in the rigid system (Fig. 4a,b) were determined by a linear fit at each *x*-coordinate, and these shifts were subtracted. Thus, the remaining shifts show the local atomic displacements due to the relaxed structure, Fig. 4i. $\Delta y$ shifts are not directly available from a single DF image when the peak $\bar{1}100$ is selected for DF imaging due to $q_y$ = 0, but can be retrieved using the available $\Delta x(x, y)$ shifts and the symmetry of the atomic distribution in the moiré structure.

For comparison, the distribution $\Delta x(x, y)$ obtained from the relaxed coordinates of twisted (0.997°) $MoS_2$ sample calculated by LAMMPS as described in ref. [25] was calculated and is shown in Fig. 4j. $\Delta x(x, y)$ from the relaxed model and the experimental DF image exhibit overall the same distribution. $\Delta x(x, y)$ recovered from the experimental DF image exhibit a slightly asymmetrical distribution which can be explained by a slight stretch of the sample and smaller values of $\Delta x(x, y)$ which can be explained by the noise and less contrast in the experimental DF image.

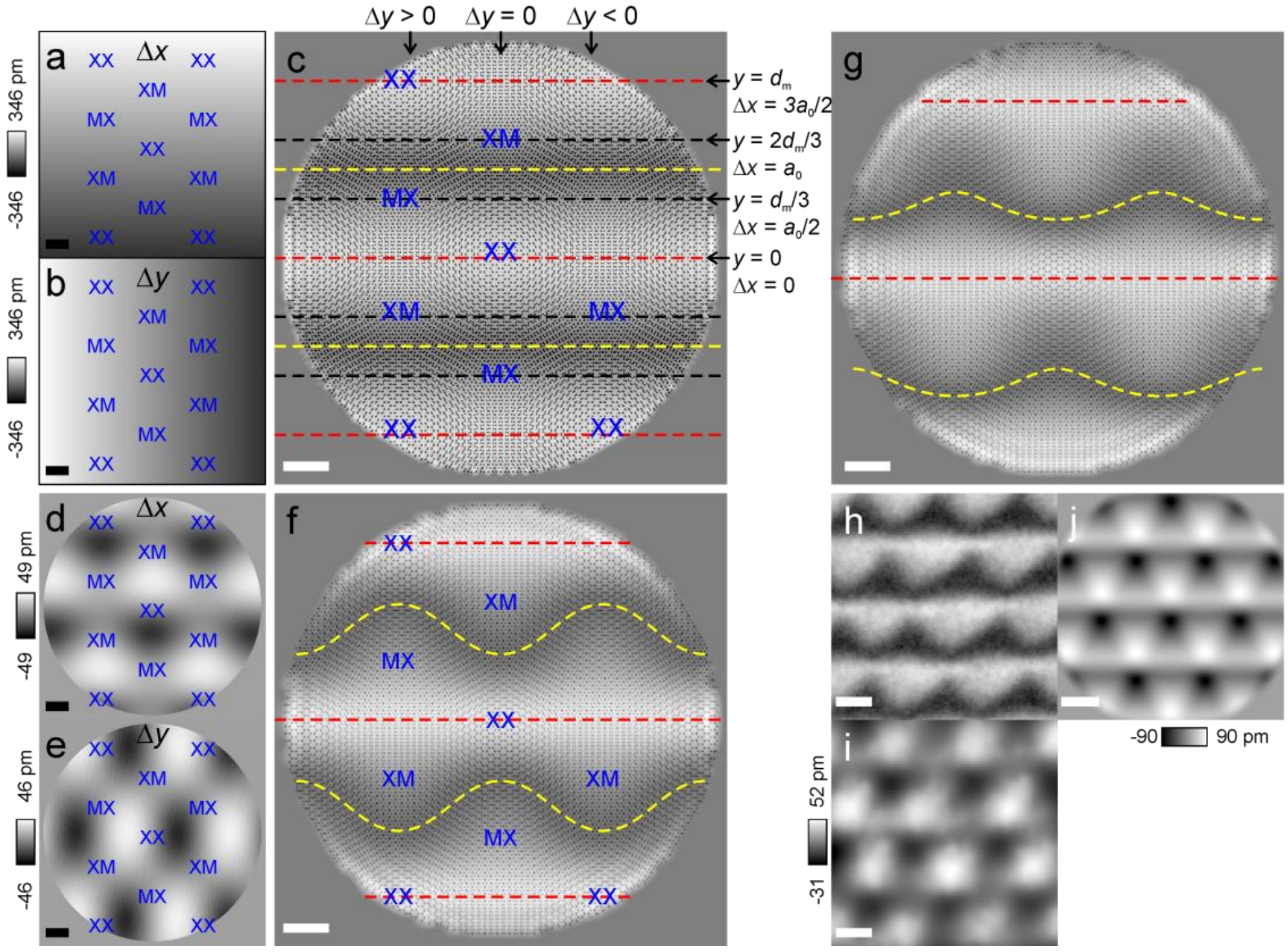

Fig. 4 Atomic coordinates and DF image contrast. (a) – (b) Simulated $\Delta x(x,y)$ and $\Delta y(x,y)$ translational shifts between two layers in a non-reconstructed (rigid) twisted (2°) bilayer system, and the corresponding (c) atomic distribution and superimposed DF image. (d) – (f) the same for the reconstructed (relaxed) system. In (d) and (e), the shifts additional to the monotone linear shifts shown in (a) and (b) are shown. (g) Atomic distribution and superimposed DF image for tilted ($\beta$ = -4°) sample. In (c), (f), and (g), the yellow dashed lines indicated the position of the minimum of the intensity, the dashed red lines – of the maximum. (h) A selected region in the experimental DF image and (i) retrieved $\Delta x(x,y)$ shifts. (j) $\Delta x(x,y)$ calculated from the relaxed coordinates of twisted (0.997°) $MoS_2$ sample using LAMMPS. Scalebar in (a) – (g) is 2 nm. Scalebar in (h) – (j) is 10 nm.

Some regions in the DF image can exhibit periodical fringes, rather than a triangular pattern (Fig. 5a and 5d). Such interference patterns can originate from two completely different phenomena: due to uniaxial strain of one of the layers or due to interference between the waves scattered by two non-relaxed layers. Here, BF imaging can help to solve the ambiguity. The BF image exhibits fringes in the case of a strained bilayer (Fig. 5b) and a uniform intensity in the case of a non-interacting bilayer sample (Fig. 5e). In addition, an intensity profile of the fringes in the DF image of a strained sample exhibits a zig-zag-like function with asymmetrical maxima (Fig. 5a), while in the case of non-interacting layers – a harmonic function (Fig. 5d). A bilayer, where one layer is strained, can create a superlattice (moiré lattice) with the lattice constant exceeding the lattice constant of a single layer by a few tens of times [26, 27]. A simulation of a twisted (2°) 2D-$MoS_2$ bilayer where the top layer is strained by a linearly increasing strain $\varepsilon_x$ = 0...0.1 over 100 nm, is shown in Fig. 5c.

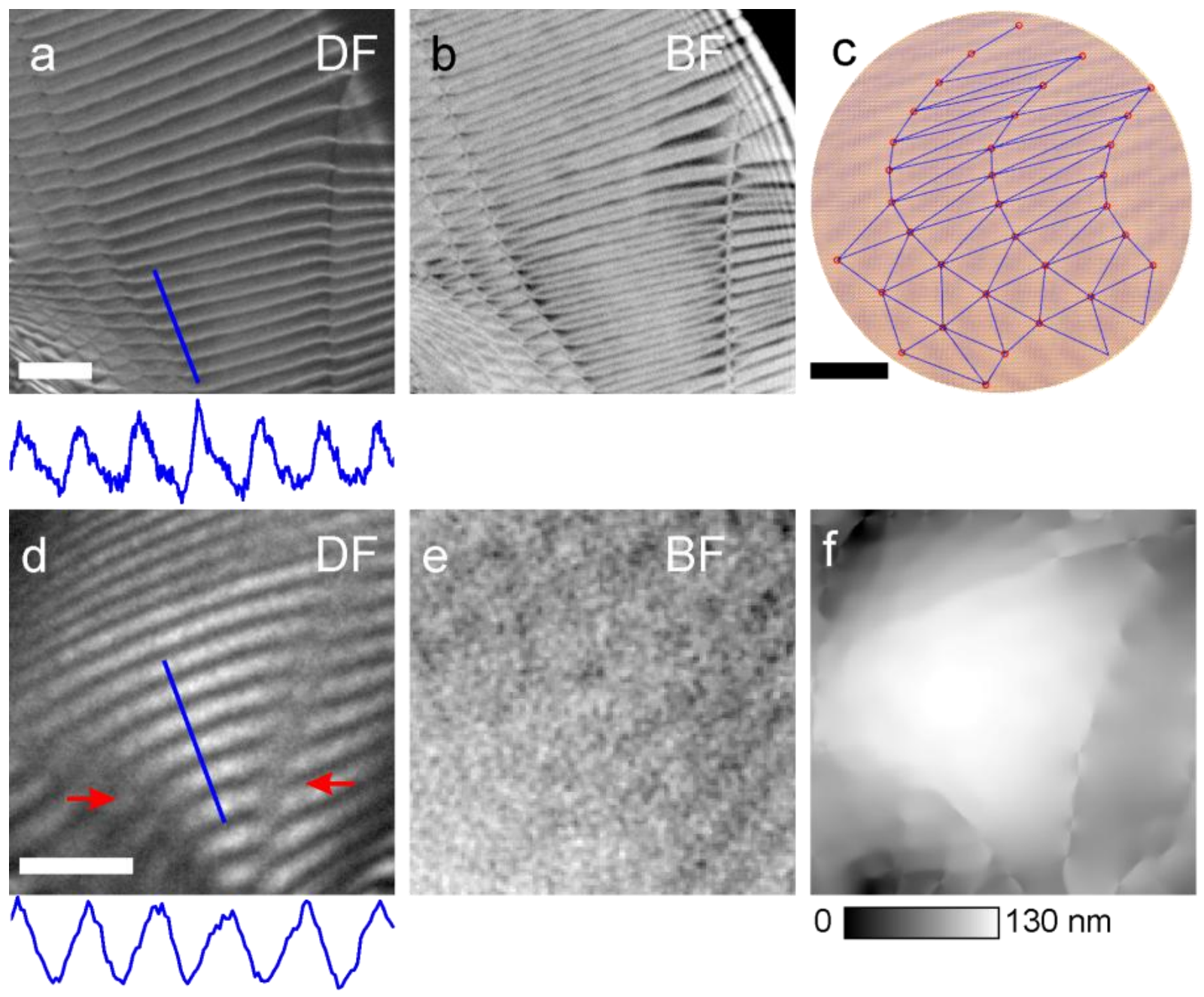

Fig. 5 DF imaging of strained 2D-$MoS_2$ bilayer. (a) DF and (b) BF images of a strained region; scalebar is 100 nm. (c) Simulated twisted (2°) 2D-$MoS_2$ bilayer, where the top layer is strained by a linearly increasing strain $\varepsilon_x$ = 0...0.1 over 100 nm. The rot circles indicate locations of the XX(P) nodes, and the blue lines indicate the domain walls observed; scalebar, 20 nm. (d) DF and (e) BF images of non-interacting layers, and the retrieved interlayer distance (f); scalebar, 20 nm. The red arrows indicate the locations where the interference pattern is shifted by pi. The curves in (a) and (d) are the intensity profiles along the blue lines.

In some regions, the distance between the layers is so large that no interaction between the two layers occurs, and the atoms are not arranged into MX or XM stacking. The DF image in this case exhibits an interference pattern formed by two waves originating from the two layers. Similar interference patterns were reported in CBED patterns of twisted bilayers, where the period and tilt of the interference fringes carry information about the interlayer distance $\Delta z$, and the interlayer distance thus can be recovered from such interference patterns [28-30]. In DF images of two non-relaxed layers, the period of the fringes is given solely by the twist angle $\theta_0$ and the lattice constant $a$: $T = a\sqrt{3}/4\sin\left(\frac{\theta_0}{2}\right)$, and does not depend on the sample tilt. Using this equation, for the period of the interference pattern in Fig. 5d of 12.7 nm, the local twist angle was evaluated to be 1.23°.

A change in the interlayer distance results in an additional phase shift between the interfering waves and thus in a shift of the entire interference pattern: $\Delta\varphi = 4\pi\lambda\Delta z / 3a^2$. A shift of the interference pattern by half of a fringe (indicated by the red arrows in Fig. 5d), that is, a phase shift by pi), corresponds to the change in the interlayer distance by about 29 nm. This also can be seen from the interlayer distribution (shown in Fig. 5f) retrieved from the interference pattern (Fig. 5d) by treating it as an off-axis hologram [31].

In conclusion, we showed that the contrast in the DF image is uniquely/unambiguously defined by the local 3D arrangement of the atoms in the sample. The interlayer distance can be retrieved from the contrast between MX and XM stacking regions provided the sample tilt is known. Within a ripple in the sample, the sample's tilt can change its sign, which for MX/XM alternating regions results in the flipping of "triangles" in the DF image. A BF image of the same region together with a DF image allows distinguishing between the sample consisting of non-interacting and strained layers. The local atomic shifts can be retrieved from the experimental DF images, which provides a novel practical tool for structural inspection in layered samples. The spatial resolution of the method is the same as that of the DF image – the size of one pixel in the DF image. Since the atomic shifts are extracted via acos of the intensity of the intensity of the DF image, their accuracy is given by the range and sampling of the intensity values and the noise level in the DF image. As the next steps, the following improvements of the method can be suggested: more precise retrieval of the 3D atomic positions could be achieved by acquiring more than two DF images of the same region by selecting different peaks in the BFP. In addition, refinement of the retrieved atomic coordinates could be done by using computational iterative methods and/or AI tools.

## Data availability statement

The data is available from corresponding author upon a reasonable request.

## Supporting Information

Additional details of simulations, sample preparation and full dark and bright field images.

## Acknowledgements

K. S. N. acknowledges support by the National Research Foundation, Singapore under its AI Singapore Programme (AISG Award No: AISG3-RP-2022-028), by the Ministry of Education, Singapore under Research Centre of Excellence award to the Institute for Functional Intelligent Materials, I-FIM (project No. EDUNC-33-18-279-V12) and by the Tier 3 program (MOE-MOET32024-0001). D.P. and T.L. acknowledge Swiss National Science Foundation (SNSF) Research Grant 200021_197107 and 10003659.

## Authors contribution

P.K. and K.S.N. prepared the samples; P.K., M.B., Z. H. and N.L. did the TEM studies, D.P. and T.L. adapted the theoretical models for relaxation, T.L. did the data analysis and simulations, T.L., K.S.N. and P.K. wrote the manuscript, T.L. and K.S.N. conceived the idea.

## Competing interests

The authors declare no competing interests.

# Supporting Information for
# Untangling 3D atomic reconstruction in twisted bilayer 2D crystals via dark field transmission electron microscopy

Pankaj Kumar[1,2], Michel Bosman[1], Nikolai Lavrentev[1], He Zheng[1], Ding Peng[3], Kostya S. Novoselov[1,2,4,], Tatiana Latychevskaia[3,*]

[1]Department of Materials Science and Engineering, National University of Singapore, Singapore, 117575, Singapore

[2]Institute for Functional Intelligent Materials, National University of Singapore, 117544 Singapore

[3]Paul Scherrer Institute, Forschungsstrasse 111, 5232 Villigen, Switzerland

[4]National Graphene Institute, University of Manchester, Oxford Road, Manchester, M13 9PL, UK

*Corresponding author: tatiana.latychevskaia@psi.ch

## Contents

## Supporting Note 1: Interlayer shifts in different stackings

| BLG | TMD(P) | | $\Delta x$ | $\Delta y$ |
|---|---|---|---|---|
| AA | XX | | $0$ | $0$ |
| AB | MX | | $-a_0$ | $0$ |
| BA | XM | | $a_0$ | $0$ |
| SP1 | DW1 | | $0$ | $\pm\frac{\sqrt{3}}{2}a_0 = \pm\frac{1}{2}a$ |
| SP2 | DW2 | | $\mp\frac{3}{4}a_0 = \mp\frac{\sqrt{3}}{4}a$ | $\pm\frac{\sqrt{3}}{4}a_0 = \pm\frac{1}{4}a$ |
| SP3 | DW3 | | $\pm\frac{3}{4}a_0 = \pm\frac{\sqrt{3}}{4}a$ | $\pm\frac{\sqrt{3}}{4}a_0 = \pm\frac{1}{4}a$ |
| IM1 | IM1 | | $\pm\frac{1}{2}a_0 = \pm\frac{1}{2\sqrt{3}}a$ | $0$ |
| IM2 | IM2 | | $\pm\frac{\sqrt{3}}{4}a_0 = \pm\frac{1}{4}a$ | $\pm\frac{1}{4}a_0 = \pm\frac{1}{4\sqrt{3}}a$ |
| IM3 | IM3 | | $\pm\frac{\sqrt{3}}{4}a_0 = \pm\frac{1}{4}a$ | $\mp\frac{1}{4}a_0 = \mp\frac{1}{4\sqrt{3}}a$ |

Here $a$ is the lattice constant and $a_0$ is the distance between the nearest atoms. Structures of BLG are shown to illustrate the relative shifts. SP – saddle point, DW – domain wall, and IM – intermediate stackings.

## Supporting Note 2: Sample tilt

Tilting the sample out of the $(x, y)$-plane along some selected axis $\alpha$ by an angle $\beta$ can be realised by the following three steps: (i) rotating the sample in the $(x, y)$-plane by $(-\alpha)$, (ii) rotating the sample in the $(x, z)$-plane by $(+\beta)$, and (iii) rotating the sample in the $(x, y)$-plane by $(+\alpha)$. The coordinates of the 3D translation vector $\Delta\vec{r}$ are transformed as $\Delta\vec{r}\,' = R_1(\alpha)R_2(\beta)R_1(-\alpha)\Delta\vec{r}$, where $R_1(\alpha)$ and $R_2(\beta)$ are the rotation matrices:

$$R_1(\alpha) = \begin{pmatrix} \cos\alpha & -\sin\alpha & 0 \\ \sin\alpha & \cos\alpha & 0 \\ 0 & 0 & 1 \end{pmatrix}, \quad R_2(\beta) = \begin{pmatrix} \cos\beta & 0 & -\sin\beta \\ 0 & 1 & 0 \\ \sin\beta & 0 & \cos\beta \end{pmatrix}.$$

## Supporting Note 3: Multislice simulations

3D coordinates of atoms in the sample $(x_i, y_i, z_i)$, where $i$ is the index of atom, are used as input. The sequence of atoms is rearranged in order of increasing $z_i$.

(i) The probing plane wave in the plane $z_1$ is calculated as

$$p_1(x, y) = 1,$$

(ii) For atom $i$=1, the transmission function at the plane $z_1$ is calculated as

$$t_1(x, y) = \exp[i\sigma v_z^{(1)}(x-x_1, y-y_1)],$$

where $\sigma$ is the interaction parameter $\sigma = 2\pi me\lambda/h^2$, $m$ is the relativistic mass of the electron, $e$ is an elementary charge, $\lambda$ is the wavelength of the electrons, $h$ is the Planck's constant, and $v_z^{(1)}(x, y)$ is the projected potential of atom (1) calculated using the parameters provided in ref. [1];

(iii) The exit wave in plane $z_1$ is calculated as

$$ew_1(x, y) = p_1(x, y)\, t_1(x, y).$$

(iv) The wave is propagated from the plane $z_1$ to the next atom 2 at plane $z_2$, using the angular spectrum method [2]. The obtained wave is the updated probing wave $ew_2$ at plane $z_2$.
(v) $i$ = 2;
(vi) the steps (i) – (v) are repeated for all atoms in the sample.

The Fourier transform of the exit wave is calculated. The obtained distribution is masked to select a first-order diffraction peak, and the filtered distribution is Fourier transformed, thus giving the wavefront in the detector plane. DF image is calculated as the squared amplitude of the wavefront in the detector plane.

## Supporting Note 4: Antiparallel (2H) 2D-$MoS_2$ system

The intensity of the signal in DF images is determined by the strength of atomic scattering amplitude. In a 2D-$MoS_2$ monolayer, one Mo atom causes the same phase shift (exhibits a similar scattering strength) as two S atoms on top of each other [1]. Thus, a 2D-$MoS_2$ layer can be approximated as a hexagonal lattice of atoms with identical scattering amplitude. In this approximation, both, a parallel (3R) and an antiparallel (2H) bilayer 2D-$MoS_2$ systems should exhibit identical intensities of DF images in different stackings. The multislice simulations show that the intensities in DF images of 3R and 2H systems are identical for small tilt angles, Supporting Figure 1.

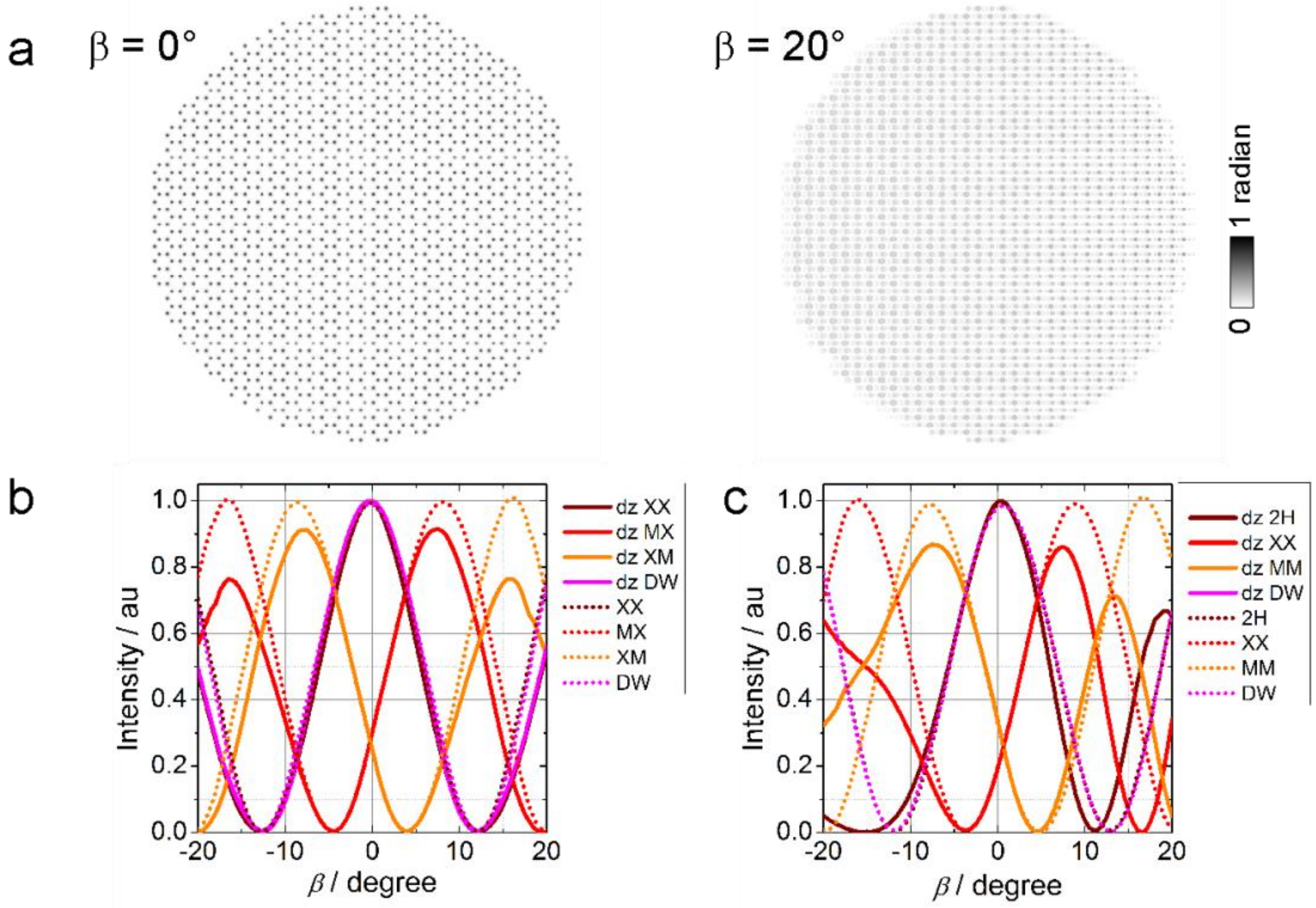

Supporting Figure 1: Simulated intensities of DF images of tilted bilayer 2D-$MoS_2$ in XX(P), MX, and XM stacking as a function of the tilt angle; simulated by multislicing for 200 keV energy electrons, $\Delta z$ = 6.2 Å. (a) Phase distributions of the exit waves at tilts $\beta$ = 0° and 20°. (b) and (c) The averaged intensity of the DF image as a function of the tilt angle $\beta$ of (b) parallel (3R) and (c) antiparallel (2H) 2D-$MoS_2$ bilayer systems; The dashed lines show the intensities when the distance between Mo and S layers within one 2D-$MoS_2$ layer was assumed to be zero.

In the simulations, the tilt angle ranged from $\beta$=-20°...+20°, at a step of 0.5°. The obtained averaged intensities of the simulated DF images are in perfect agreement with those shown in Fig. 2g,h, when the atoms within one 2D-$MoS_2$ layer are assumed to be in one plane (that is, neglecting the distance between Mo and S atomic layers). In more realistic situation, when the layers of S atoms are separated by $\Delta z$ = ± 154 pm from the layer of Mo atoms, the matching is achieved only for small tilts in the range of -5°...+5°, Supporting Figure 1b,c.

Regarding an analytical solution for an antiparallel system: In the case of parallel stacking, the bilayer system can be represented as two identical layers with a relative translational shift $\Delta\vec{r}$, and an analytical expression for the intensity of the DF image can be easily derived (Eq.(1)). In the case of antiparallel stacking, the system cannot be represented as two identical layers with a relative translational shift. However, it is possible to represent each layer (a hexagonal lattice) as a set of two trigonal lattices. The overall atomic distribution in an antiparallel system can then be represented as four trigonal lattices that are translationally shifted relative to each other, and an analytical equation similar to Eq. (1), in principle, can be obtained.

## Supporting Note 5: Sample preparation

The steps for TEM sample preparation are as follows and also shown in Supporting Figure 2.

1. 2D-$MoS_2$ is exfoliated on a clean polydimethylsiloxane (PDMS), and its thickness is identified using optical contrast. Prior to exfoliation, the PDMS substrate was cleaned with exposure to UV-ozone for 12 minutes. It is critical to clean the PDMS; otherwise, the organic residues from PDMS will form a thin layer of polymer on the flake. Once a suitable flake is found, the PDMS stamp is positioned over a SiNx TEM grid (SiNx membrane with a 2.5 um array of holes). To enhance the adhesion between the membrane and the flake, a thin layer of Ti/Au (2/3 nm) is pre-deposited.

2. The PDMS with the 2D-$MoS_2$ flake is brought into contact with the gold-coated TEM SiN membrane, ensuring that only half of the flake contacts the membrane. We then slowly lift the PDMS stamp leaving the part of the flake that was in contact with the membrane. The stronger adhesion of 2D-$MoS_2$ with the gold-coated membrane allows the flake to be split in half.

3. Next, we rotate the stage with the desired twist angle. We then position the PDMS stamp over the flake on the membrane and lower the arm to make contact. We release the remaining half of 2D-$MoS_2$ flake on the existing 2D-$MoS_2$ flake on the membrane.

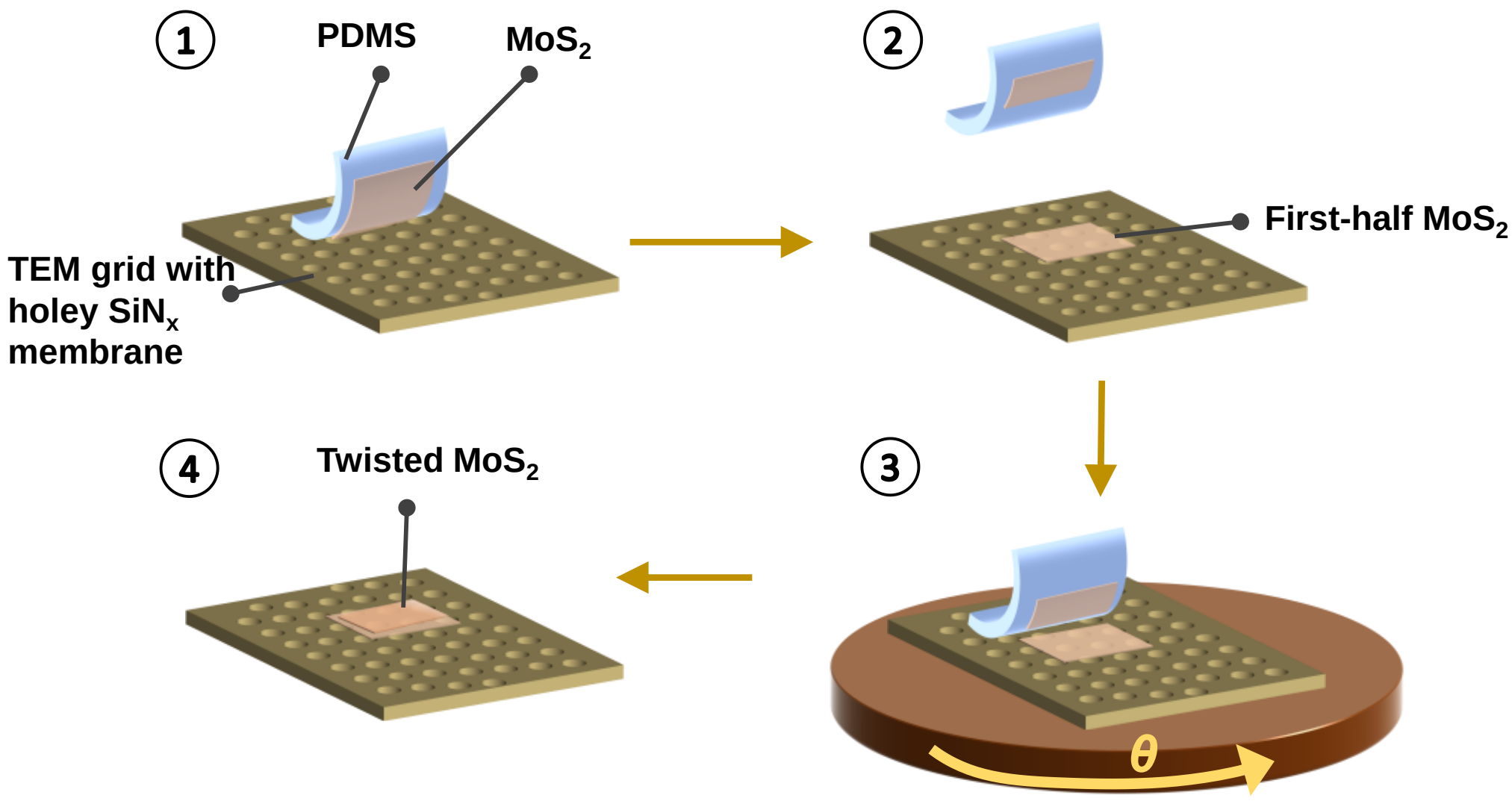

Supporting Figure 2: Schematic diagram of PDMS based dry transfer method to fabricate twisted bilayer 2D-$MoS_2$ sample for TEM study.

No solvent was involved in the entire process to minimize contamination. Contrary to the widely used hBN pickup method wherein a thin hBN is used as a supporting layer to prepare TEM samples [3], no hBN support layer was used. The obtained twisted bilayer 2D-$MoS_2$ was freely suspended over holes in the SiN membrane. The sample was imaged with 200 keV electrons using a Titan TEM (Thermo Fisher Scientific) without aberration-corrector, and a Gatan OneView camera for BF and DF imaging. A first-order diffraction peak was selected for DF imaging. The sample was slightly tilted to achieve the maximal contrast.

## Supporting Note 6: Experimental DF and BF images

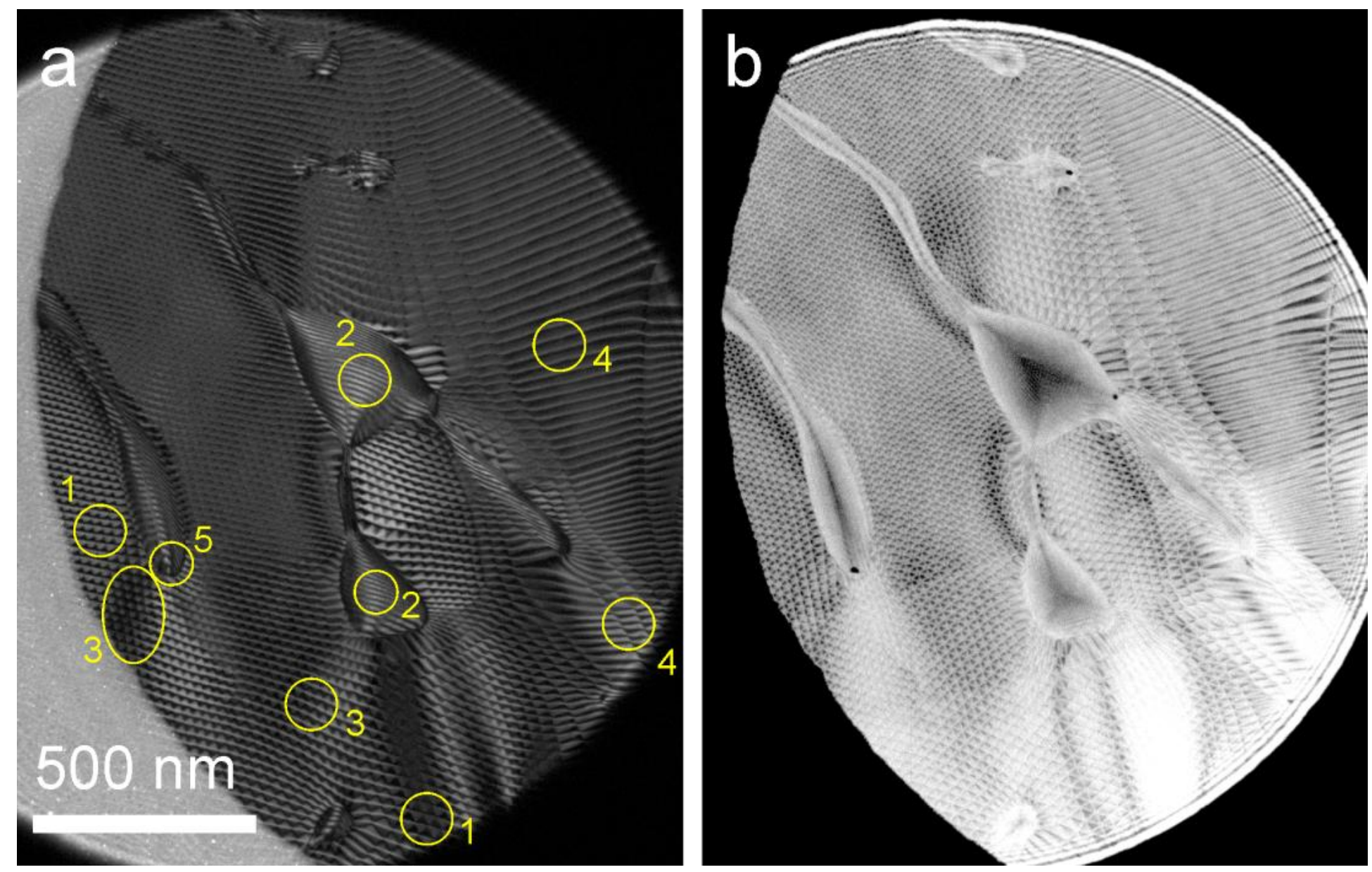

Supporting Figure 3: Experimental (a) dark field and (b) bright field images of 2D-$MoS_2$. The characteristic regions are indicated: (1) MX/XM stacking regions arranged into "triangles", (2) non-interacting layers, (3) contrast inversion due to sample tilt, and (4) strained regions.